\documentclass[conference]{IEEEtran}

\usepackage{cite}
\input epsf
\usepackage{graphicx}
\usepackage{ tipa }
\usepackage{amsmath}
\usepackage{algorithm}  
\usepackage{algorithmicx}  
\usepackage{algpseudocode}  
\usepackage{cases}
\usepackage{subfigure}
\usepackage{array}
\usepackage{color}
\usepackage{soul}
\usepackage{amssymb}

\usepackage{url}

\begin{document}
%
\title{ Potential Integral Equations in Electromagnetics}

\author{\IEEEauthorblockN{Jie Li, Balasubramaniam Shanker}
\IEEEauthorblockA{Department of Electrical and Computer Engineering\\
Michigan State University, East Lansing, MI 48824}
\and
\IEEEauthorblockN{Xin Fu}
\IEEEauthorblockA{Department of Electronic and Electrical Engineering\\
The University of Hong Kong, Pokfulam, Hong Kong, China\\}
}

\maketitle

\begin{abstract}
In this work, a new integral equation (IE) based formulation is proposed using vector and scalar potentials for electromagnetic scattering. The new integral equations feature decoupled vector and scalar potentials that satisfy Lorentz gauge. The decoupling  of the two potentials allows low-frequency stability. The formulation presented also results in Fredholm integral equations of second kind. The spectral properties of second kind integral operators leads to a well-conditioned system. 
\end{abstract}

\IEEEpeerreviewmaketitle

\section{Introduction}

Recently two similar formulations \cite{Vico2016,Liu2015} have been proposed to solve the low frequency breakdown based on the auxiliary vector and scalar potentials instead of electromagnetic fields. In \cite{Vico2016}, two fully decoupled integral equations (IEs) are formulated for vector potential ($\mathbf{A}$) and scalar potential ($\phi$) respectively. The operators in those IEs are defined to result in IE formulations of the second kind, which is significantly useful in finite element analysis. 
Liu et al. \cite{Liu2015} with $\mathbf{A}$ based integral equation also shows the low-frequency stability.  In both papers, the vector potential and scalar potential are related through Lorentz gauge. 
However the second approach does not guarantee Fredholm integral equations of the second kind. What's more, at low frequency (approaching to zero), the system equation is nearly a saddle point problem that needs special care. In this work,  starting from the rigorously derived representation theorem for $\mathbf{A}$ \cite{Chew2014}, a new decoupled potential integral equation method is formulated for more generalized dielectric problems.
The proposed formulation features stability and is comprised of Fredholm integral equations of the second kind. Detailed  discussion including spectral analysis and verification will be presented at the conference. This paper lays out the basic formulation together with ideas and tools for spectral analysis.

\section{Decoupled potential Formulation}
In this section, decoupled $\mathbf{A}-\phi$ descriptions of scattering from the homogeneous body will be presented. The formulation will be consistent with Lorentz gauge. In this work, time harmonic factor $e^{j\omega t}$ is assumed and suppressed.
\subsection{ Decoupled Potential Representations and Boundary Conditions}
Vector potential $\mathbf{A}$ and scalar potential $\phi$ satisfy vector and scalar Helmholtz equations, respectively. The following representations for scattered potential can be derived; 
\begin{equation}
\label{equ:rep_vec}
\begin{aligned}
& \mathbf{A}^{sca}=  \mathcal{S}_k[\hat{n}'\times \nabla' \times \mathbf{A}(\mathbf{r}')] + \nabla \times \mathcal{S}_k [ \hat{n}'\times \mathbf{A}(\mathbf{r}')]  \\
  & ~~~~~-  \mathcal{S}_k[ \hat{n}'  \nabla' \cdot \mathbf{A}(\mathbf{r}')]  - \nabla \mathcal{S}_k[ \hat{n}'  \cdot \mathbf{A}(\mathbf{r}')]
  \equiv \mathbf{\mathcal{L}}^{\bf A}_k[\{ {\bf A} \}]
\end{aligned}
\end{equation}
\begin{equation}
\label{equ:rep_sca}
\phi^{sca} = -\mathcal{S}_k[\frac{\partial \phi (\mathbf{r}')}{\partial n'}] + \mathcal{D}_k[\phi(\mathbf{r}')]
\equiv \mathcal{L}^{\phi}_k[\{ \phi \}]
\end{equation}
where $\mathcal{S}_k$ and $\mathcal{D}_k$ are single and double layer potential operators respectively.

Electric and magnetic fields can be decomposed into ${\bf A}-\phi$ forms (${\bf E} = -j\omega {\bf A} - \nabla \phi$ and 
${\bf H} = \frac{\nabla \times {\bf A}}{\mu} $).
From the following conditions of electric and magnetic fields,
\begin{equation}
\begin{aligned}
\hat{n} \times \mathbf{E}_1 & = \hat{n} \times \mathbf{E}_2   
,~~ \hat{n} \times \mathbf{H}_1 = \hat{n} \times \mathbf{H}_2   \\
\epsilon_1 \hat{n} \cdot \mathbf{E}_1 &=  \epsilon_2 \hat{n} \cdot \mathbf{E}_2   
~~\text{and}~~ \mu_1 \hat{n} \cdot \mathbf{H}_1 =\mu_2\hat{n} \cdot \mathbf{H}_2   ，
\end{aligned}
\end{equation}
one can set up two sets of stronger conditions for the two decoupled potentials. For scalar potential, one has
\begin{equation}
\label{equ:bc_die_phi}
\begin{aligned}
\phi_1   = \phi_2 + V  
~~\text{and}~~ \epsilon_1 \hat{n} \cdot \nabla  \phi_1  = \epsilon_2  \hat{n} \cdot  \nabla \phi_2  ,
\end{aligned}
\end{equation}
where $V$ is a jump term allowed in scalar potential, and each separated object corresponds to one constant value.
For vector potential, one has
\begin{equation}
\label{equ:bc_die_a}
\begin{aligned}
& \hat{n} \times \mathbf{A}_1  =  \hat{n} \times \mathbf{A}_2, ~~~
\epsilon_1 \hat{n} \cdot  \mathbf{A}_1  =  \epsilon_2  \hat{n} \cdot  \mathbf{A}_2 \\
~~& \text{and}~~\dfrac{1}{\mu_1}\hat{n}\times \nabla \times \mathbf{A}_1  = \dfrac{1}{\mu_2} \hat{n}\times \nabla \times \mathbf{A}_2  .   
\end{aligned}
\end{equation}

Due to the two facts that (1) $\mathbf{A}$ and $\phi$ have to satisfy the Lorentz gauge and (2) the charge neutrality requirement should be imposed if not satisfied implicitly, one can get additional conditions for $\phi$ and ${\bf A}$ respectively: (1)
$\int \frac{\partial \phi_1  }{\partial n} dS' = 0 $ for $\phi$ and (2) 
$\nabla \cdot  \mathbf{A}_1  =  \nabla \cdot  \mathbf{A}_2  + C $
and $ \int \hat{n} \cdot  \mathbf{A}_1  dS' =  0  $ for ${\bf A}$.
$C$ is a jump term in divergence of $\mathbf{A}$ as in $\phi$. It's worth noting that two jump terms could be set to zero to derive even stronger boundary conditions. In each case, one can denote the application of boundary conditions on the interface as $\{ \phi_2\} = \mathcal{B}(\{ \phi_1\},V)$ or $\{ {\bf A}_2 \} = {\bf \mathcal{B}}(\{ {\bf A}_1\},C)$.
\subsection{Well-conditioned Integral Equations}
Given the representation theorems and the boundary conditions, one can construct integral equations for dielectrics. Without loss of generality, one
can set the jump terms to zero and represent the total $\phi$ and $\partial_n \phi$ just outside the interface as
\begin{subequations}
\begin{align}
\phi_{i} &= \phi_{inc}\delta_{1i} + \mathcal{L}^{\phi}_{k_i}[\{\phi_{i}\}]   \\
\partial_n \phi_{i}
&= \partial_n \phi_{inc} \delta_{1i}
+ \partial_n\mathcal{L}^{\phi}_{k_i}[\{\phi_{i}\}]
\end{align}
\end{subequations}
where $\delta_{11}=1$ and $\delta_{12}=0$, and the operand $\{ \phi_{i} \}$ denotes all the possible components required as in Eq.\ref{equ:rep_sca} to represent the scattered scalar potential. Linear combination of $\phi_1$ and $\phi_2$ will give us one IE with four unknowns associated with fields just outside and inside of the interface, as does the linear combination of their normal derivative terms. Applying boundary conditions given earlier would reduce those unknowns into two unknowns (associated with outside of the interface) and the jump (associated with each separated body). Therefore, on the interface, one has
\begin{subequations}
\begin{align}
& (1+\alpha )\phi_{1} - X[\phi_1, \alpha]  = \phi_{inc} \\
& \dfrac{\beta\epsilon_1 + \epsilon_2}{\epsilon_2}\frac{\partial \phi_{1}}{\partial n} 
- \frac{\partial}{\partial n} X[\phi_1, \beta]=\{ \phi_{1} \}] = \frac{\partial \phi_{inc} }{\partial n}
\end{align}
\end{subequations}
where $X[\phi_1, \sigma]=  \mathcal{L}^{\phi}_{k_1} [\{ \phi_{1} \}] +\sigma \mathcal{L}^{\phi}_{k_2}[\mathcal{B}(\{ \phi_{1}\})]$ with $\sigma$ being $\alpha$ or $\beta$.

After rewriting the IEs with explicit single and double layer operators, one can get a system with its diagonal being identity operator plus compact operators. To obtain a second kind system, one needs to handle the hypersingular term carefully. Furthermore, note that there are many choices factors $\alpha$ and $\beta$, but only specific ones can give nice spectral property of the operators. To avoid the system of the first kind, $\beta$ has to be $-1$ to generate the weakly singular integral operator.

Following a similar procedure, one can get the vector potential integral equations. By denoting necessary components of  ${\bf A}$ with 
$\{ {\bf A}\} = [\hat{n}\times \nabla \times {\bf A} , \hat{n}\times  {\bf A}, \nabla \cdot {\bf A}, \hat{n} \cdot {\bf A}  ]^T 
=  [{\bf A}^{v1} , {\bf A}^{v2}, A^{s1}, A^{s2}]^T $, one can get the following vector potential integral equations.
\begin{subequations}
\label{equ:IE_A}
\begin{align}
& (1+\alpha) {\bf A}_1^{v1}
-  \hat{n}\times \nabla \times  {\bf X}[{\bf A}_1, \alpha]  =  {\bf A}_{inc}^{v1}\\
&(1+\beta) {\bf A}_1^{v2}
-  \hat{n}\times  {\bf X}[{\bf A}_1, \beta]   =  {\bf A}_{inc}^{v2}\\
&(1+\gamma) A_1^{s1}
-  \nabla \cdot  {\bf X}[{\bf A}_1, \gamma]   =  A_{inc}^{s1}\\
&(1+\delta) A_1^{s2}
-  \hat{n}\cdot  {\bf X}[{\bf A}_1, \delta]   =  A_{inc}^{s2}
\end{align}
\end{subequations}
where ${\bf X}[{\bf A}_1, \sigma] = \mathcal{L}^{{\bf A}}_{k_1}[\{ {\bf A}_{1} \}] 
+ \sigma \mathcal{L}^{{\bf A}}_{k_2}[\mathcal{B}(\{  {\bf A}_1 \})] $ with $\sigma$ being $\alpha, \beta, \gamma$ or $\delta$. Similarly as in scalar potential case, specific linear factors $\sigma$ should be chosen to cancel the singularity beyond $\frac{1}{R}$ to produce compact off-diagonal operators. This is very similar to the situation in M\"{u}ller formulation. Then one can also show that this IE set is also of the second kind, different from the extension of \cite{Liu2015} in \cite{Liu2016}.

It's noted that  the charge neutrality mentioned earlier should be imposed if the discretization scheme cannot guarantee that.

\section{Analysis on the Sphere}
In this section, the integral system is solved on a PEC sphere of radius $r=a$ for a special case where the scalar potential vanishes \cite{Chew2014,Liu2015}. 
Those two unknown components of $\mathbf{A}$ for PEC case are ${\bf A}^{v1}$ and $A^{s2}$ and hence only two equations out of Eqs.\ref{equ:IE_A}. There are several possible choices that will give very different behaviors in terms of eigen properties and conditioning at low frequency. It can be shown that the first one and the last one should be chosen to lead to second kind IEs. Other reduced forms include the scheme used in \cite{Liu2015}, which is not optimal. 

The unknown quantities can be represented by vector and scalar spherical harmonics as in \cite{Li2015}.
Following \cite{Li2015}, one can easily show that the mode orthogonality still holds in the discrete system. Hence the system is block-diagonal, with each of the block (corresponding to mode $(n,m)$) being $3 \times 3$ in size. 
Analytic derivations lead to only four nonzero matrix entry in each block $ [Z^{nm}_{ij}]$ for mode $(n,m)$. The nonzero entries can be evaluated analytically (given as follows), which allows analytic inversion of the system. 
\begin{subequations}
\begin{align}
\begin{split}
Z_{11}^{nm}& 
=  a^2 + ja^2  \mathbb{Z}_{n}^{(4)}(ka)  \mathbb{Z}_{n}^{(1)'}(ka) ]  
\end{split}  \\
 \begin{split}
Z_{22}^{nm}& 
= a^2 -  ja^2   \mathbb{Z}_{n}^{(1)}(ka)  \mathbb{Z}_{n}^{(4)'}(ka) ] 
 \end{split} \\
 \begin{split}
Z_{31}^{nm}& 
=  \dfrac{a^2\sqrt{n(n+1)}}{jk}\Big[ z_n^{(4)}(ka) \mathbb{Z}_n^{(1)'}(ka)  \\
& ~~~~~~~~~~~~~~~~~~~~~ + ka z_n^{(4)'}(ka)z_n^{(1)}(ka)\Big]
 \end{split}\\
  \begin{split}
Z_{33}^{nm}& 
= a^2 -jk^2a^4  {z^{(4)}}'_n(ka) z^{(1)}_n(ka)  
 \end{split}
\end{align}
\end{subequations}
where $z^{(1)}(\cdot)$ and $z^{(4)}(\cdot)$ are spherical Bessel function and spherical Hankel function of the second kind respectively, and $\mathbb{Z}^{(1)}(\cdot)$ and $\mathbb{Z}^{(4)}(\cdot)$ are spherical Riccati Bessel function and spherical Riccati Hankel function of the second kind respectively. 

Together with the canonical incidence  $\mathbf{A}_{inc}$, one can get the analytic solution of the scattering problem. Given plane wave incidence, one can construct the corresponding $\mathbf{A}-\phi$ representation, which are again represented by using incoming spherical vector and scalar wave functions. Results have been obtained for the above IEs,  matching exactly those from Mie series approach.

More importantly, given the analytic matrix entries, the explicit "impedance" matrix can be used to show the spectral property of the integral equations derived in the previous section. Detailed analysis will be presented at the conference.

\section{Conclusion}
Starting from the representation theorems for both scalar and vector potentials,
one can derive a decoupled and the second kind Fredholm integral equation based formulation for electromagnetic scattering. The new IE features frequency stability and spectral properties that are essential to solve the linear system, especially when the iterative method is used. Note that this work is not immune from interior resonance at higher frequency for the reduced PEC case. Similar techniques as in combined field integral equation and decoupled potential IE in \cite{Vico2016} are expected to remove the resonance in existing operators.


\bibliographystyle{IEEEtran}
\bibliography{dpie}

\end{document}